\documentclass[twocolumn,showpacs,preprintnumbers,superscriptaddress,amsmath,amssymb,aps,pre]{revtex4}
\usepackage{graphicx}% Include figure files
\usepackage{dcolumn}% Align table columns on decimal point
\usepackage{bm}% bold math
\usepackage{color}% color

\begin{document}
% \begin{CJK*}{GBK}{Song} % Use default fonts from CJK (see below)

%\preprint{RCE Working Paper No. 2008-05}

\title{Universal and nonuniversal allometric scaling behaviors in the visibility graphs of world stock market indices}

\author{Meng-Cen Qian}
 \affiliation{School of Management, Fudan University, Shanghai 200433, China} %
\author{Zhi-Qiang Jiang}
 \affiliation{School of Business, East China University of Science and Technology, Shanghai 200237, China} %
 \affiliation{School of Science, East China University of Science and Technology, Shanghai 200237, China} %
 \affiliation{Research Center for Econophysics, East China University of Science and Technology, Shanghai 200237, China} %
\author{Wei-Xing Zhou}
 \email{wxzhou@ecust.edu.cn}
 \affiliation{School of Business, East China University of Science and Technology, Shanghai 200237, China} %
 \affiliation{School of Science, East China University of Science and Technology, Shanghai 200237, China} %
 \affiliation{Research Center for Econophysics, East China University of Science and Technology, Shanghai 200237, China} %
 \affiliation{Engineering Research Center of Process Systems Engineering (Ministry of Education), East China University of Science and Technology, Shanghai 200237, China} %
 \affiliation{Research Center on Fictitious Economics \& Data Science, Chinese Academy of Sciences, Beijing 100080, China}%

\date{\today}

\begin{abstract}
The investigations of financial markets from a complex network perspective have unveiled many phenomenological properties, in which the majority of these studies map the financial markets into one complex network. In this work, we investigate 30 world stock market indices through their visibility graphs by adopting the visibility algorithm to convert each single stock index into one visibility graph. A universal allometric scaling law is uncovered in the minimal spanning trees, whose scaling exponent is independent of the stock market and the length of the stock index. In contrast, the maximal spanning trees and the random spanning trees do not exhibit universal allometric scaling behaviors. There are marked discrepancies in the allometric scaling behaviors between the stock indices and the Brownian motions. Using surrogate time series, we find that these discrepancies are caused by the fat-tailedness of the return distribution, the nonlinear long-term correlation, and a coupling effect between these two influence factors.
\end{abstract}

\pacs{05.45.Tp, 05.40.-a, 05.45.Df, 89.75.Da, 89.65.Gh}

%\pacs{87.23.Ge}{Dynamics of social systems} %
%\pacs{89.65.-s}{Social and economic systems} %
%\pacs{89.75.-k}{Complex systems} %

\maketitle

% \end{CJK*}

\section{Introduction}
\label{S1:Intro}

Econophysics is an interdisciplinary field which adopts ideas, tools and theories from statistical mechanics, nonlinear science, complexity science and applied mathematics to understand the emerging complexity and self-organized macroscopic behaviors of economic systems, with special interest paid to financial markets \cite{Mantegna-Stanley-2000,Bouchaud-Potters-2000,Sornette-2003,Malevergne-Sornette-2006,Zhou-2007}. Econophysists are particularly interested in unveiling different universal behaviors in financial markets \cite{Vandewalle-Boveroux-Minguet-Ausloos-1998-PA,Stanley-Amaral-Gopikrishnan-Plerou-2000-PA,Zhou-Yuan-2005-PA,Plerou-Stanley-2007-PRE,Plerou-Stanley-2008-PRE,Stanley-Plerou-Gabaix-2008-PA}, following the phenomenological framework \cite{Bouchaud-2008-Nature,Lux-Westerhoff-2009-NP}.

In recent years, complex network theory has witnessed a flourishing progress \cite{Watts-Strogatz-1998-Nature,Barabasi-Albert-1999-Science,Albert-Barabasi-2002-RMP,Newman-2003-SIAMR,Dorogovtsev-Mendes-2003,Boccaletti-Latora-Moreno-Chavez-Hwang-2006-PR}.
It is natural that a wealth of studies have been carried out from a complex network perspective. The current economic crisis calls for a deeper understanding of the dynamics of economic activities on the global economic network \cite{Schweitzer-Fagiolo-Sornette-VegaRedondo-Vespignani-White-2009-Science}. The studies in this field can be classified into two types based on how the network is constructed. The studies of the first type deal with many time series to form a complex network with each node standing for a time series and the weight of a link between two nodes characterized by the correlation coefficient of the two time series
\cite{Laloux-Cizean-Bouchaud-Potters-1999-PRL,Plerou-Gopikrishnan-Rosenow-Amaral-Stanley-1999-PRL,Plerou-Gopikrishnan-Rosenow-Amaral-Guhr-Stanley-2002-PRE} or by the distance between the two time series \cite{Mantegna-1999-EPJB,Onnela-Chakraborti-Kaski-Kertesz-2002-EPJB,Onnela-Chakraborti-Kaski-Kertesz-2003-PA}.

Concerning the studies of the second type, different mapping methods have been proposed to convert time series into different kinds of networks, including cycle networks based on the local extrema and their distance in the phase space \cite{Zhang-Small-2006-PRL,Zhang-Sun-Luo-Zhang-Nakamura-Small-2008-PD}, segment correlation networks \cite{Yang-Yang-2008-PA,Gao-Jin-2009-PRE}, nearest neighbor networks \cite{Xu-Zhang-Small1-2008-PNAS}, n-tuple networks based on the fluctuation patterns \cite{Li-Wang-2006-CSB,Li-Wang-2007-PA}, the visibility of nodes \cite{Lacasa-Luque-Ballesteros-Luque-Nuno-2008-PNAS}, space state networks based on conformational fluctuations \cite{Li-Yang-Komatsuzak-2008-PNAS}, bin transition networks \cite{Shirazi-Jafari-Davoudi-Peinke-Tabar-Sahimi-2009-JSM}, and recurrence networks \cite{Marwan-Donges-Zou-Donner-Kurths-2009-XXX,Donner-Zou-Donges-Marwan-Kurths-2009-XXX}. The visibility algorithm has been diversely used to investigate stock market indices \cite{Ni-Jiang-Zhou-2009-PLA}, human strive intervals \cite{Lacasa-Luque-Luque-Nuno-2009-EPL}, occurrence of hurricanes in the United States \cite{Elsner-Jagger-Fogarty-2009-GRL}, foreign exchange rates \cite{Yang-Wang-Yang-Mang-2009-PA}, and energy dissipation rates in three-dimensional fully developed turbulence \cite{Liu-Zhou-Yuan-2009-XXX}.

In this work, we study the allometric scaling behavior of spanning trees extracted from the visibility graphs of 30 stock market indices all over the world. Both universal and nonuniversal scaling behaviors are reported, which is reminiscent of the universal scaling behavior of the weighted world trade networks \cite{Duan-2007-EPJB}. The paper is organized as follows. In Sec. \ref{S1:Method}, we describe briefly the methodology adopted. Section \ref{S1:Empirics} presents the empirical findings. We study the impact of the length of the stock indices on the allometric behaviors in Sec.~\ref{S1:FSE}. Section \ref{S1:factors} explores the driving factors that cause the different behaviors between the stock indices and the Brownian motions. And Section \ref{S1:Conclusion} concludes.

\section{Methodology}
\label{S1:Method}

Here we briefly explain the methodology used in this study. For each stock market index, a unique visibility graph is constructed. A maximal spanning tree (MaxST), a minimal spanning tree (MinST), and 100 random spanning trees (RanSTs) are extracted from the constructed visibility graph. For each spanning tree, an allometric scaling analysis is carried out and a scaling exponent is determined.

\subsection{Construction of visibility graph}

Consider the price series $p(t)$ of a stock market index with a length of $N$. We can transform the time series into complex networks by applying the  visibility algorithm \cite{Lacasa-Luque-Ballesteros-Luque-Nuno-2008-PNAS}. Each data point in the time series is regarded as a node in the complex network, and an edge is drew connecting two nodes according to the rule that the two corresponding data points can see each other in the diagram of the time series. Mathematically, two arbitrary data points $p(t_i)$ and $p(t_j)$ have visibility if any other data point $p(t_k)$ located between them fulfills
\begin{equation}
\frac{p(t_j)-p(t_k)}{t_j-t_k} > \frac{p(t_j)-p(t_i)}{t_j-t_i}.
 \label{Eq:VG}
\end{equation}
We assign the average acceleration of price movement, defined as $[\ln p(t_j) - \ln p(t_i)]/(t_j-t_i)$, as the weight of the edge $[t_i,t_j]$, where $t_i<t_j$.

\subsection{Construction of spanning trees}

We extract three different kinds of spanning trees (MaxST, MinST, RanST) from each visibility graph. The algorithms for building MaxST, MinST, and RanST are given in the following.
\begin{description}
  \item[MaxST:] First, select an edge with a maximum weight as
  the first edge of the MaxST. Second, select a new edge with a
  maximum weight among the edges which connect the tree and make
  sure that no loop is introduced. Third, repeat the second step till
  all nodes are added into the tree.
  \item[MinST:] First, select an edge with a minimum weight as
  the first edge of the MinST. Second, select a new edge with a
  minimum weight among the edges which connect the tree and make
  sure that no loop is introduced. Third, repeat the second step till
  all nodes are added into the tree.
  \item[RanST:] First, arbitrarily select an edge as the first edge of the RanST.
  Second, randomly select a new edge among the edges which connect the tree but
  make sure that no loop is introduced.
  Third, repeat the second step till all nodes are added into the tree.
\end{description}

\subsection{Allometric scaling}

Allometric scaling laws are ubiquitous in complex systems evolving on complex networks, such as the metabolism of organisms and ecosystems river networks \cite{West-Brown-Enquist-1997-Science,Enquist-Brown-West-1998-Nature,West-Brown-Enquist-1999-Science,Enquist-West-Charnov-Brown-1999-Nature,Banavar-Maritan-Rinaldo-1999-Nature,Enquist-Economo-Huxman-Allen-Ignace-Gillooly-2003-Nature}, the food webs \cite{Garlaschelli-Caldarelli-Pietronero-2003-Nature}, the world trade webs \cite{Duan-2007-EPJB}, the world investment networks \cite{Song-Jiang-Zhou-2009-PA}, and so forth. The original model of the allometric scaling on a spanning tree was developed by Banavar, Maritan, and Rinaldo \cite{Banavar-Maritan-Rinaldo-1999-Nature}. The node with the maximum degree is considered as the root of a spanning tree. Each node of a spanning tree is assigned a number 1, and two values $A_i$ and $C_i$ are defined for each node
$i$ in an iterative manner as follows:
\begin{equation}
  A_i = \sum_j A_j + 1~~{\rm{and}}~~ C_i = \sum_j C_j + A_i,
  \label{Eq:AC}
\end{equation}
in which $j$ stands for all the nodes linked {\it{from}} $i$ \cite{Banavar-Maritan-Rinaldo-1999-Nature}. The allometric scaling relation is then highlighted by the power law relation between $C_i$ and $A_i$:
\begin{equation}
 C \sim A^{\eta},
 \label{Eq:AC:eta}
\end{equation}
where the leaf nodes with $A=C=1$ should be excluded from the estimation of the scaling exponent $\eta$ \cite{Garlaschelli-Caldarelli-Pietronero-2003-Nature}.

Any spanning tree can range in principle between two extremes, that is, the chain-like trees and the star-like trees. For chain-like trees we have $\eta=2^-$, while for start-like trees we have $\eta = 1^+$. Therefore, $1< \eta < 2$ for all spanning trees. It should be note that not all trees exhibit such an allometric scaling behavior, for instance the classic Cayley trees \cite{Jiang-Zhou-Xu-Yuan-2007-AICHEJ}.

\section{Universal scaling in the visibility graphs constructed from stock market indices}
\label{S1:Empirics}

The data sets we analyzed contain 30 stock market indices all over the world, which are retrieved from Yahoo! Finance at http://finance.yahoo.com. A list of the index names is given below together with the abbreviations, countries or areas and starting dates of the time series used for analysis in the ensuing parentheses: Amsterdam Exchange Index (AEX, Netherlands, 3 January 2000), ATX Vienna (ATX, Austria, 11 November 1992), Euronext BEL-20 (BFX, Belgium, 11 February 2005), BSE Sensex (BSESN, India, 1 July 1997), Ibovespa (BVSP, Brazil, 27 April 1993), CMA GENL Index (CMA, Egypt, 26 May 2003), Dow Jones Industrial Average (DJIA, USA, 1 October 1928), CAC 40 Index (FCHI, France, 1 March 1990), FTSE 100 Index (FTSE, UK, 2 April 1984), DAX Index (GDAXI, Germany, 26 November 1990), S\&P/TSX Composite Index (GSPTSE, Canada, 3 January 2000), Hang Seng Index (HSI, Hong Kong, 31 December 1996), Jakarta Composite Index (JKSE, Indonesia, 1 July 1997), FTSE Bursa Malasia KLCI Index (KLSE, Malaysia, 3 December 1993), Korea Composite Stock Price Index (KOSPI, Korea, 1 July 1997), Merval Buenos Aires (MERV, Argentina, 8 October 1996), MIBTEL Index (MIBTEL, Italy, 3 January 2000), MXX IPC (MXX, Mexico, 1 November 1991), NIKKEI 225 (N225, Japan, 4 January 1984), NASDAQ Composite (NASDAQ, USA, 2 February 1971), NZX 50 Index (Gross) (NZ50, New Zealand, 30 April 2004), OMXS All Share Index (OMXSPI, Sweden, 8 January 2001), Oslo Exchange All Share Index (OSEAX, Norway, 7 February 2001), IGBM (SMSI, Spain, 2 January 2002), Standard and Poor's 500 Index (S\&P500, USA, 3 January 1950), Shanghai Stock Exchange Composite Index (SSEC, China, 4 January 2000), Swiss Market Index (SMI, Switzerland, 9 November 1990), Straits Times Index (STI, Singapore, 28 December 1987), Tel Aviv TA-100 Index (TA100, Israel, 1 July 1997), and Taiwan Stock Exchange Corporation Weighted Index (TWII, Taiwan, 2 July 1997). The ending dates of all the indices are 25 August 2009.

\begin{figure}[htb]
\centering
\includegraphics[width=7cm]{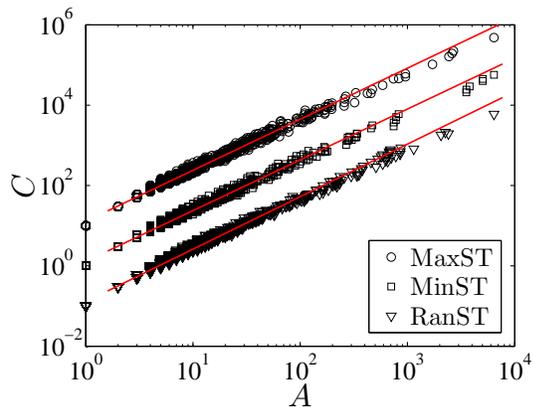}
\caption{\label{Fig:VG:MST:AS:FTSE} Allometric scaling behavior of spanning trees extracted from the visibility graph of the FTSE 100 Index. The data points for MinST and RanST are transformed vertically by a factor of 10 and 0.1 for better visibility. The solid lines are the best power-law fits.}
\end{figure}

For each stock index, a visibility graph is constructed and its maximum spanning tree (MaxST) and minimum spanning tree (MinST) are determined uniquely. In addition, 100 random spanning trees (RanSTs) are also derived from the visibility graph. For each tree, an allometric analysis is carried out and the two sequences of $A$ and $C$ are calculated. Figure \ref{Fig:VG:MST:AS:FTSE} shows the allometric scaling behaviors of the MaxST, the MinST and a randomly selected RanST associated with the FSTE 100 Index. Nice power-law relationships are observed between $A$ and $C$. A linear regression to the data finds that $\eta = 1.271 \pm 0.002$ for the MaxST, $\eta = 1.264 \pm 0.002$ for the MinST, and $\eta = 1.308 \pm 0.002$ for the RanST, respectively.

We find that all the spanning trees exhibit excellent allometric scaling behaviors for all the indices. The corresponding power-law exponents for different indices are reported in table~\ref{Tb:eta}. We find that the allometric scaling exponents of the MaxSTs for different indices are very close to each other, centered around $1.281 \pm 0.009$ (mean $\pm$ std). The situation for the MinSTs is the same as the MaxSTs, in which the power-law exponents fluctuate slightly around $1.274 \pm 0.011$. In contrast, the power-law exponents of the RanSTs are basically larger than 1.3 except for DJIA, NASDAQ and S\&P 500, which are significantly larger than the exponents of both the MaxSTs and the MinSTs. We infer that the appearance of smaller values of $\eta_{\rm{RanST}}$ for DJIA, NASDAQ, and S\&P 500 is due to the fact that these three indices have much more data points compared with other indices.

\begin{table}[htp]
 \caption{\label{Tb:eta} Allometric scaling exponents $\eta$ of spanning trees (MaxST, MinST, and RanST) for different indices.}
 \medskip
 \centering
 \begin{tabular}{cccc}
 \hline \hline
  stock index & $\eta_{\rm{MaxST}}$ & $\eta_{\rm{MinST}}$ & $\eta_{\rm{RanST}}$ \\
  \hline
  AEX & $1.276(3)$ & $1.278(3)$ & $1.329(3)$ \\
  ATX & $1.282(3)$ & $1.270(3)$ & $1.312(2)$ \\
  BFX & $1.270(5)$ & $1.269(4)$ & $1.349(4)$ \\
  BSESN & $1.288(3)$ & $1.281(3)$ & $1.310(3)$ \\
  BVSP & $1.299(3)$ & $1.290(3)$ & $1.315(2)$ \\
  CMA & $1.277(12)$ & $1.250(8)$ & $1.315(7)$ \\
  DJIA & $1.284(1)$ & $1.267(1)$ & $1.274(1)$ \\
  FCHI & $1.280(2)$ & $1.269(2)$ & $1.313(2)$ \\
  FTSE & $1.271(2)$ & $1.264(2)$ & $1.308(2)$ \\
  GDAXI & $1.284(2)$ & $1.270(2)$ & $1.312(2)$ \\
  GSPTSE & $1.273(3)$ & $1.260(4)$ & $1.333(3)$ \\
  HSI & $1.276(2)$ & $1.276(2)$ & $1.304(2)$ \\
  JKSE & $1.284(3)$ & $1.280(3)$ & $1.309(3)$ \\
  KLSE & $1.276(3)$ & $1.284(3)$ & $1.318(2)$ \\
  KOSPI & $1.279(3)$ & $1.292(3)$ & $1.307(3)$ \\
  MERV & $1.278(3)$ & $1.282(3)$ & $1.318(2)$ \\
  MIBTEL & $1.274(3)$ & $1.274(3)$ & $1.327(3)$ \\
  MXX & $1.281(3)$ & $1.280(3)$ & $1.312(2)$ \\
  N225 & $1.276(2)$ & $1.271(2)$ & $1.308(2)$ \\
  NASDAQ & $1.308(2)$ & $1.267(2)$ & $1.271(2)$ \\
  NZ50 & $1.260(4)$ & $1.276(4)$ & $1.348(4)$ \\
  OMXSPI & $1.278(3)$ & $1.283(4)$ & $1.313(3)$ \\
  OSEAX & $1.278(3)$ & $1.266(4)$ & $1.309(3)$ \\
  SMSI & $1.297(5)$ & $1.294(5)$ & $1.345(5)$ \\
  SP500 & $1.280(1)$ & $1.256(1)$ & $1.282(1)$ \\
  SSEC & $1.285(4)$ & $1.264(4)$ & $1.313(3)$ \\
  SMI & $1.279(2)$ & $1.267(2)$ & $1.313(2)$ \\
  STI & $1.273(2)$ & $1.276(2)$ & $1.317(2)$ \\
  TA100 & $1.290(3)$ & $1.286(3)$ & $1.324(3)$ \\
  TWII & $1.291(3)$ & $1.285(3)$ & $1.325(3)$ \\
  \hline \hline
 \end{tabular}
\end{table}

\section{Finite-size effect}
\label{S1:FSE}

Table \ref{Tb:eta} shows that the allometric scaling exponents for a given type of spanning tree are very close to each other. However, there are still fluctuations around the corresponding average values. It is possible that the exponent is dependent on the length of the index time series. It is thus necessary to further investigate if there is a finite-size effect, which is crucial to the validation of universality.

\subsection{The case of MaxST and MinST}

\begin{figure}[htb]
\centering
\includegraphics[width=7cm]{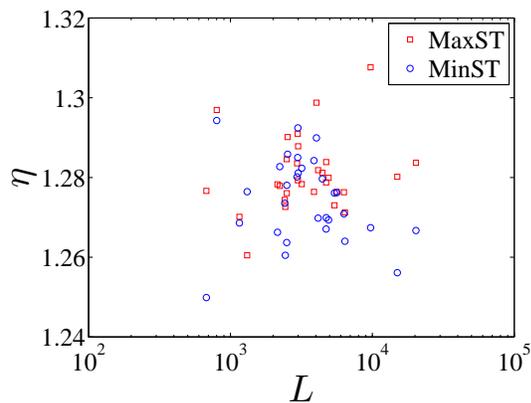}
\caption{\label{Fig:VG:MST:eta:L:MST} (Color online) Dependence of the allometric scaling exponents $\eta_{\rm{MaxST}}$ and $\eta_{\rm{MinST}}$ on the length $L$ of stock indices. No evident finite-size effect is observed.}
\end{figure}

Figure \ref{Fig:VG:MST:eta:L:MST} illustrates the dependence of $\eta_{\rm{MaxST}}$ and $\eta_{\rm{MinST}}$ as a function of the stock index length $L$. No evident trend is identified by eye-balling. We fit the data for MaxSTs and MinSTs using a linear model
\begin{equation}
 \eta_{\rm{MaxST,MinST}} = a + b L.
 \label{Eq:VG:MST:eta:L:MST}
\end{equation}
For the MaxSTs, $a=1.279$ and $b=3.949\times10^{-7}$ and the corresponding $p$-values from the Student's t-test are 0 and 0.355, respectively.
For the MinSTs, $a=1.278$ and $b=-8.315\times10^{-7}$ and the corresponding $p$-values are 0 and 0.082, respectively. If we regress the exponent against $\ln{L}$, the $p$-value of $b$ is 0.322 for the MaxSTs and 0.314 for the MinSTs. It is clear that the coefficient $b$ is identical to 0 and the exponents $\eta_{\rm{MaxST}}$ and $\eta_{\rm{MinST}}$ are independent of $L$.

\subsection{The case of RanST}

\begin{figure}[htb]
\centering
\includegraphics[width=7cm]{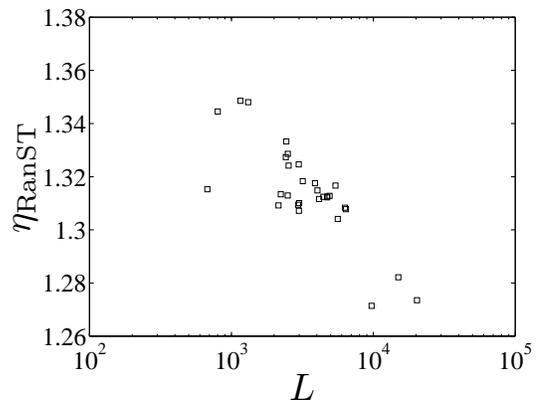}
\caption{\label{Fig:VG:MST:eta:L:RST} Dependence of the allometric scaling exponent $\eta_{\rm{RanST}}$ on the length $L$ of stock indices. An evident finite-size effect is observed.}
\end{figure}

For the RanSTs extracted from the visibility graphs of stock indices, the three allometric scaling exponents $\eta_{\rm{RanST}}$ of the US market indices (DJIA, S\&P 500, NASDAQ) are significantly less than the others, which is a signal of the possible presence of finite-size effect.
Figure \ref{Fig:VG:MST:eta:L:RST} plots $\eta_{\rm{RanST}}$ as a function of $L$ for all the indices. It is evident that $\eta_{\rm{RanST}}$ decreases logarithmically with the increase of index length:
\begin{equation}
 \eta_{\rm{RanST}} = a + b \ln{L},
 \label{Eq:VG:MST:eta:L:RST}
\end{equation}
where $a=1.471$ and $b=-0.019$, both of which are significantly different from 0 with the $p$-values less than $10^{-6}$.

\subsection{Numerical tests}

\begin{figure}[htb]
\centering
\includegraphics[width=7cm]{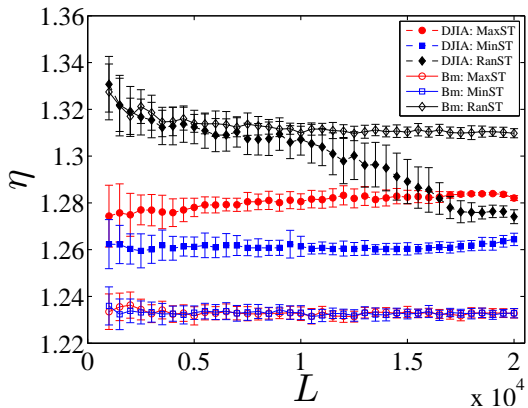}
\caption{\label{Fig:VG:MST:FiniteSize} (Color online) Dependence of the averaged exponent $\eta$ with respect to the length $L$ for the DJIA and the Brownian motions.}
\end{figure}

In order to further investigate the impact of the time series length on the allometric scaling exponents of the three types of spanning trees, we design and conduct two numerical tests. We take subseries with different lengths from the longest index DJIA to perform the allometric scaling analysis. For each length $L$, 100 subseries are randomly extracted and the corresponding scaling exponents are calculated from their MaxSTs, MinSTs, and RanSTs. Figure \ref{Fig:VG:MST:FiniteSize} illustrates the dependence of the averaged exponent $\eta$ with respect to the length $L$ for the DJIA subseries. A linear regression of $\eta_{\rm{MinST}}$ against $L$ using Eq.~(\ref{Eq:VG:MST:eta:L:MST}) gives that $a=1.261$ and $b=4.093\times10^{-8}$ with the $p$-values being 0 and 0.16, respectively. It means that $\eta_{\rm{MinST}}$ is independent of $L$. For the MaxSTs, we find that $\eta_{\rm{MaxST}}$ increases with $L$ linearly
\begin{equation}
 \eta_{\rm{MaxST}} = 1.276  + 4.555\times10^{-7} L,
 \label{Eq:VG:MaxST:FSE}
\end{equation}
where the linear coefficients are statistically significantly different from zero with both the $p$-values being zero. In contrast, $\eta_{\rm{RanST}}$ exhibits an evident decreasing trend. A linear regression gives $a=1.327$ and $b=-2.581\times10^{-6}$, whose $p$-values are both nulls. These observations are consistent with the results in Table \ref{Tb:eta}. In addition, we find that $\eta_{\rm{MinST}}<\eta_{\rm{MaxST}}$ for all $L$'s, while $\eta_{\rm{RanST}}$ becomes less than $\eta_{\rm{MaxST}}$ when $L>16500$ and less than $\eta_{\rm{MinST}}$ for much greater $L$.

For comparison, we synthesize Brownian motions with different lengths. For each length, 100 time series are generated and the allometric scaling exponents are determined. The results are also illustrated in Fig.~\ref{Fig:VG:MST:FiniteSize}. The scaling exponents $\eta_{\rm{MaxST}}$ and $\eta_{\rm{MinST}}$ are found to be independent of $L$ since linear regressions show that the $p$-values of the corresponding $b$'s are 0.17 and 0.06, respectively. On the other hand, the scaling exponent $\eta_{\rm{MinST}}$ exhibits a decreasing trend for small $L$ and then reaches a constant. For all $L$'s, $\eta_{\rm{MinST}}\approx\eta_{\rm{MaxST}}<\eta_{\rm{RanST}}$ for the Brownian motions.

We stress that there are marked discrepancies between the scaling exponents of the MinSTs (and the MaxSTs as well) of the DJIA and Brownian motions. It calls for an investigation of influence factors on the allometric behaviors of the MinSTs and MaxSTs of the stock market indices. Note that a comparison between the behaviors of the stock indices and the Brownian motions is not out of blue. Indeed, stock prices are assumed to follow a geometric Brownian motion in the celebrated Black-Scholes model of option pricing \cite{Black-Scholes-1973-JPE}.

\section{Influence factors on the allometric scaling behavior}
\label{S1:factors}

To understand the marked discrepancies between the allometric behaviors of the stock market indices and the Brownian motions, further numerical experiments are needed. Comparing with Brownian motions, we find there are three potential factors that may have influence on the behavior of any time series, that is, the fat-tailedness of the probability distribution, the linear long-term correlation, and the nonlinear long-term correlation \cite{Zhou-2009-EPL}. It is well established that there is no linear long-term correlation in the returns of stock indices. Hence, two factors remain for further investigations. For completeness, we first study the impact of linear long-term correlations based on fractional Brownian motions.

\subsection{Fractional Brownian motions}

We generate fractional Brownian motions through wavelet transform with the Hurst indexes varying from 0.05 to 0.95. We obtain 100 realizations for each Hurst index, and each realization has 5000 data points.

In the case of Brwonian motions where the Hurst index $H = 0.5$, remarkable power-law behaviors are observed between $A$ and $C$ for the three spanning trees. The least square linear fit to the data for each spanning tree yields an estimate of the power-law exponent, which results in $\eta = 1.233 \pm 0.004$ for the MaxST, $\eta = 1.233 \pm 0.004$ for the MinST, and $\eta = 1.314 \pm 0.004$ for the RanST. Note that the node $(1,1)$ is excluded in the implementation of fitting. The plots of $A$ with respect to $C$ for the other Hurst indexes share the similar pattern as the plots for $H = 0.5$, but the power-law exponents are different.

\begin{figure}[htb]
\centering
\includegraphics[width=7cm]{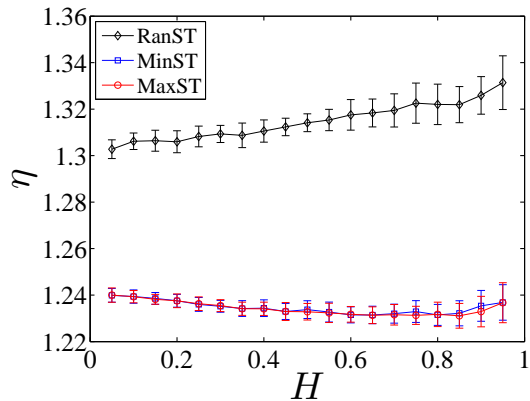}
\caption{\label{Fig:MST:fBm:Heta} Plots of the dependence of allometric scaling exponent $\eta$ on the Hurst index $H$ for three spanning trees.}
\end{figure}

Figure~\ref{Fig:MST:fBm:Heta} shows the allometric scaling exponent $\eta$ as a function of the Hurst index $H$ for three different spanning trees. For the MaxST and the MinST, the two curves almost overlap onto a same curve. Furthermore, both curves show a linear decreasing trend
\begin{equation}
 \eta = a + b H,
 \label{Eq:Heta}
\end{equation}
where $a=1.238$ and $b=-0.008$ for the MaxSTs and $a=1.238$ and $b=-0.007$ for the MinSTs with all the $p$-values less than 0.01. The variation of these two scaling exponents is actually very slight and all the exponents are embedded in the interval $[1.23,1.24]$. In contrast, the power-law scaling exponent $\eta$ for the RanSTs increases with the Hurst index $H$, which can also be modeled by Eq.(\ref{Eq:Heta}). A linear regression finds that $a=1.301$ and $b = 0.027$ with the associated $p$-values being zero.

\subsection{Origin of the difference of allometric scaling exponents between financial series and Brownian motions}

Compared with the Brownian motions, the financial series exhibit fat-tailed PDF and nonlinearity as well \cite{Zhou-2009-EPL}. We design several tests to verify the contribution of the two factors on the allometric scaling behavior. In the tests, different surrogate data are used. There is no linear long-term correlation in the surrogate data. We use the FTSE index as an example, and all the surrogate time series have the same length as the FTSE data.

The surrogates of the first type (termed ``Surr 1'') have the same probability distribution of returns as the FTSE index, but without any nonlinearity. We can shuffle the original the corresponding return series $r(t)$ to remove the nonlinearity, and conduct cumulative summation to reconstruct the prices. Alternatively, we can determine the empirical distribution of the FTSE returns and generate surrogate return series \cite{Press-Teukolsky-Vetterling-Flannery-1996}. We find that the surrogate time series obtained from either methods give the same scaling exponents.

The surrogates of the second type (termed ``Surr 2'') preserve the nonlinearity of the FTSE data but with Gaussian return distributions. We generate a Brownian motion, rearrange its increment series to ensure that the resulting series has the same rank ordering as the FTSE returns \cite{Bogachev-Eichner-Bunde-2007-PRL,Zhou-2008-PRE,Zhou-2009-EPL}.

The surrogates of the third type (termed ``Surr 3'') preserve both the nonlinearity and the probability distribution of the original FTSE returns. We synthesize a surrogate return series based on the empirical distribution of the FTSE returns and then rearrange the rank orders of the data to introduce nonlinear correlations.

For each test, 100 surrogates are generated, and the allometric scaling exponents are determined for the corresponding MaxSTs, MinSTs, and RanSTs. The results are presented in Table \ref{TB:Tests}, which are compared with the FTSE index and the Brownian motions. First of all, the scaling exponents of the random spanning trees are close to each other for the five types of data. For the MaxSTs and MinSTs, we have
\begin{equation}
 \eta_{\rm{Bm}} < \eta_{\rm{Surr~1}},\eta_{\rm{Surr~2}} < \eta_{\rm{Surr~3}} = \eta_{\rm{FTSE}}.
 \label{Eq:VG:MST:Tests}
\end{equation}
Therefore, both the nonlinearity and the fat-tailedness have influence on the allometric scaling. However, only one factor is not capable of explaining the difference of $\eta$ between the FTSE index and the Brownian motions. A coupling effect between the two factors also has contribution.

\begin{table}[htb]
  \centering
  \caption{Determining the influence of the fat-tailedness in the probability and the nonlinear long-term correlation on the allometric scaling behavior of the FTSE index.}\label{TB:Tests}
  \medskip
  \begin{tabular}{ccccccccccccccccccccc}
    \hline\hline
    Test & MaxST & MinST & RanST \\
    \hline
FTSE      & $ 1.271 \pm 0.002 $ & $ 1.264 \pm 0.002 $ & $ 1.308 \pm 0.002 $ \\
Surr 1    & $ 1.263 \pm 0.004 $ & $ 1.250 \pm 0.005 $ & $ 1.306 \pm 0.005 $ \\
Surr 2    & $ 1.255 \pm 0.002 $ & $ 1.255 \pm 0.002 $ & $ 1.315 \pm 0.004 $ \\
Surr 3    & $ 1.272 \pm 0.002 $ & $ 1.262 \pm 0.003 $ & $ 1.310 \pm 0.004 $ \\
Bm        & $ 1.241 \pm 0.002 $ & $ 1.243 \pm 0.002 $ & $ 1.303 \pm 0.004 $ \\
    \hline\hline
  \end{tabular}
\end{table}

\section{Conclusion}
\label{S1:Conclusion}

In summary, we have investigated the allometric scaling behaviors of the maximal spanning trees, the minimal spanning trees and the random spanning trees of the visibility graphs constructed from 30 worldwide stock indices with different sizes. All the spanning trees exhibit nice allometric scaling behaviors. We found that the average scaling exponent is $\eta_{\rm{MaxST}}=1.281\pm0.009$ for the maximal spanning trees and $\eta_{\rm{MinST}}=1.274\pm0.011$ for the minimal spanning trees. Using subseries extracted from the longest index series of the DJIA, we found that the exponent for the minimal spanning trees is independent of the length of the index, while the exponent for the maximal spanning trees exhibits a weak increasing trend. In contrast, the exponent for the random spanning trees decreases fast with the increase of the length. Therefore, the visibility graphs of the world stock market indices exhibit both universal and nonuniversal allometric scaling behaviors.

Numerical simulations showed that Brownian motions with different sizes exhibit different allometric behaviors in the associated maximal, minimal, and random spanning trees. Surrogate time series with different features were synthesized to study the origin of the discrepancy between Brownian motions and stock indices. We conclude that the fat-tailedness of the return probability distribution, the nonlinear long-term correlation, and a coupling effect between them can explain exactly the discrepancy.

\begin{acknowledgments}
This work was partially supported by the National Natural Science
Foundation of China (Grant No. 10905023), the Shanghai Educational Development
Foundation (Grant Nos. 2008CG37 and 2008SG29), and the Program for New Century
Excellent Talents in University (Grant No. NCET-07-0288).
\end{acknowledgments}

\bibliography{/home/zqjiang/Science/Papers/Auxiliary/Bibliography}
%\bibliography{E:/Papers/Auxiliary/Bibliography}

\end{document}